# Clearing the Path for Software Sustainability

Jennifer Gross
Dept. Information Technology
Uppsala University
Uppsala, Sweden
jennifer.gross@it.uu.se

Sofia Ouhbi
Dept. Information Technology
Uppsala University
Uppsala, Sweden
sofia.ouhbi@it.uu.se

## ABSTRACT
The advancement of software sustainability encounters notable challenges, underscoring the necessity for understanding these challenges to facilitate significant progress and pave the way for effective solutions to advance software sustainability. This paper outlines key challenges identified in literature based on findings from a tertiary study. Challenges identified include: confusion regarding the definition of software sustainability, uncertainty about when to consider sustainability in software development, lack of assessment metrics and tools, narrow perspectives on sustainability in software systems, insufficient awareness and education, and a lack of serious considerations in practice. The paper aims at clarifying the confusion surrounding software sustainability to motivate effective solutions. The provided recommendations aim to give a more organized approach towards advancing sustainable software development, emphasizing comprehensive strategies, the integration of sustainability as a fundamental aspect of software development, actionable research directions, and the cultivation of a common understanding of sustainable software.

## 1 INTRODUCTION

In recent years, there has been a growing emphasis on software sustainability due to society's increasing reliance on information and communication technology (ICT) solutions. Despite receiving increasing attention from academia and industry, investigations into sustainable software systems are still in their infancy and are facing several challenges that need to be tackled [8, 40, 50–52]. Software sustainability is a complex, multifaceted concept encompassing different dimensions of sustainability. Some researchers [11, 49] connect software sustainability with the dimensions outlined for software development in the Brundtland report [9]: economic, social, and environmental. Other researchers, such as in the Karlskrona Manifesto [5], link it to two additional dimensions: individual and technical.

Within the research community, there is no consensus on the definition of sustainable software [41]. The absence of a clear definition and standardized methods for software sustainability hinders the advancement of the field. The complexity of software sustainability lies also in its categorization as sustainability-in, sustainability-by, or both. Sustainability-in, or green-in, refers to software developed with sustainability as a primary goal, while sustainability-by, or green-by, refers to software used to achieve environmental sustainability through external actions [26]. Sustainability can also be considered in terms of its effects, both positive and negative [6]. First-order effects are the direct impacts of the production, use, and disposal of ICTs. Second-order effects are the indirect impacts of the ICTs. Third-order effects are the broader consequences that these changes may have on other systems, potentially leading to global effects. For example, creating software that establishes a secure system can enhance global trade.

The pursuit of sustainability in software systems often falls short due to several challenges facing the field. In this paper, we aim to understand the key challenges hindering the progress of software sustainability to pave the way for effective sustainable software solutions. To do so, we conducted a tertiary study to analyze literature reviews published in the last five years to include recent results from other reviews in the field. The primary studies analyzed were selected from Scopus, one of the largest databases of peer-reviewed research publications. By synthesizing diverse challenges and limitations extracted from the literature, we identified key challenges facing the field of software sustainability. We then provided a list of recommendations for researchers and practitioners to pave the way for effective sustainable software solutions.

## 2 METHOD

To gain deeper insights into the challenges and limitations of software sustainability as documented in existing literature, we conducted a tertiary study focusing on literature reviews in the field. A tertiary study is a specific type of review, characterized as a review of reviews [36]. The search for literature reviews was carried out at the beginning of April 2024, utilizing one of the largest databases, Scopus. Scopus indexes peer-reviewed papers from venues that adhere to specific quality criteria and are also indexed in other digital libraries such as IEEE, ACM Digital, Science Direct, and SpringerLink. Our study is driven by the following research question: *"What are the challenges and limitations facing software sustainability as reported in the literature?"*

With this research question in mind, we developed the following search string to identify potential papers: **Within Article Title**: software AND (sustainability OR sustainable OR green*) **AND within Article Title, Abstract, Keywords**: "literature review" OR "mapping study" OR review OR snowballing

We applied the following exclusion criteria (EC) to filter out the candidate studies:

- EC1. Studies not focused on software sustainability.
- EC2. Studies that do not qualify as literature reviews.
- EC3. Notes, editorials, introductions to proceedings, and other document types that are not relevant to the study.

A total of 98 documents were initially retrieved, covering the period from 1985 to 2024. To concentrate on recent reviews that address relevant challenges, we narrowed down the search results to the last five years, from 2019 to 2024 (EC1). This is to ensure the inclusion of recent results from other reviews in the field. This refinement yielded a total of 57 documents, all of which were in English. From Scopus, we extracted a CSV file[1] containing all relevant data, including citation information (e.g., document title, year, source title, and DOI) and data related to abstracts and keywords. Through the application of EC by reviewing the titles and abstracts of the candidate studies, we excluded 27 documents based on EC1, 9 documents based on EC2, and 3 documents based on EC3. A total of 18 papers were selected for this tertiary study.

## 3 RESULTS

The results of the tertiary study are presented in Table 1 and Table 2. Of the 18 studies included in this tertiary study, 9 selected studies were published in journals, including three published in Sustainability journal, 8 studies were conference papers, with two studies each from the International Conference on Electrical Engineering and Informatics, and one review was published as a chapter book.

## 4 CHALLENGES

Below is a compilation of the challenges and limitations identified as a result of this tertiary study.

### 4.1 Confusion

The discourse surrounding software sustainability often finds itself mired in confusion, due to a lack of a clear understanding of the term itself [4, 25, 32]. The term *sustainability* is frequently used without a clear delineation of its implications, resulting in ambiguity in its application [38]. Defined in the dictionary as "to endure," sustainability, if narrowly interpreted, could be equated to durability, a notion that is quite limiting [50].

Often used without a concrete specification of what aspects are being considered [38], software sustainability, in its broader context, transcends software projects and has an impact on society and the environment. The term "software sustainability" was used by Seacord et al. in 2003 to refer to a team's ability to update software based on customer needs and implement these changes [46]. However, as time progressed, the term evolved to encompass a wide array of sustainability dimensions, ranging from three to five dimensions. As highlighted by Venters et al. [50], while numerous approaches have been proposed, a clear definition of software sustainability remains elusive, rendering the dimensional perspective still murky [24, 41]. Adding to the complexity are the varied approaches taken by different research groups, such as the focus on energy evaluation [24], or the software product durability [41].

The inclusion of concepts like sustainable-in and sustainable-by, as well as green-in and green-by, in the terminology of sustainable software contributes to the confusion [12, 15]. Not commonly introduced in this manner, these terms can lead to misunderstandings, particularly when referring to "sustainable software." Being aware of such categorization of software either in or by, a person might be wondering when hearing "sustainable software" whether it is denoting software that is inherently sustainable (sustainable-in), or software that aids in making another system sustainable (sustainable-by). "Sustainability-by" or "green-by," originally introduced to bridge the gap left by "green-in," encompasses other fields like computational sustainability and sustainable-HCI, providing a framework to discuss the second-order effects of software sustainability [26].

Moreover, the delineation between software itself and its broader effects remains unclear. Software and its effects are interdependent, so focusing solely on the software and disregarding its use or societal/environmental/economic impact limits our understanding of its true needs [27]. In essence, the challenge of confusion in software sustainability literature arises from the lack of a precise and commonly accepted definition. The evolving nature of the term, coupled with diverse interpretations and terminology, contributes to the complexity of the discourse.

### 4.2 Uncertainty about when to consider sustainability in software development

Navigating the incorporation of sustainability into software development poses a significant challenge due to the varied perspectives on when and how to introduce sustainable practices. The uncertainty surrounding the incorporation of sustainability into software development processes reflects the evolving nature of the field.

Karita et al. [32] underscores the importance of early engagement of stakeholders and the integration of sustainability throughout the software development lifecycle. This call is echoed by Ibrahim et al. [29] highlighting the absence of sustainability integration or waste management models in development processes. The landscape becomes more complex with the discovery of inadequate software engineering practices hindering sustainability efforts [42, 47]. Swache et al.'s [48] identification of the absence of validation or verification mechanisms in existing models adds to this uncertainty.

While aspects of software sustainability are already embedded within software quality assessments [2, 14, 16, 22, 23, 34], the trajectory for incorporating broader sustainability evaluations remains unclear. Calero et al., for instance, have expanded the ISO/IEC 25010 model to include an environmental dimension, emphasizing its concurrent consideration with other (sub-)characteristics during evaluations [49]. While certain characteristics of software quality, like maintainability and usability, find acknowledgment within existing sustainability models, other sustainability aspects transcending the technical dimension remain underrepresented.

---

[1] CSV available here: https://shorturl.at/ezNO3

**Table 1: Tertiary study results Part I**

| Ref. | Year | Title | Challenges or limitations reported in the selected publication |
|---|---|---|---|
| [47] | 2023 | A Systems Thinking Approach to Improve Sustainability in Software Engineering—A Grounded Capability Maturity Framework | Current research is limited by its narrow focus on specific areas like the environment, energy, or a single phase of software development. |
| [4] | 2023 | Requirements engineering for sustainable software systems: a systematic mapping study | The study highlights a lack of consensus on sustainability in requirements engineering, with no approaches tailored to agile settings. Though some strategies were tested in industrial case studies, none were widely adopted. This poses a challenge in maturing these approaches for practitioner use, further challenged by the absence of supporting software tools for sustainable software system requirements. |
| [50] | 2023 | Sustainable software engineering: Reflections on advances in research and practice | The literature identifies challenges in software architecture sustainability and metrics. Key points are the need for better methods to maintain architectural consistency, limitations in current techniques that focus on code over architecture, challenges in architectural reconstruction, uncertainty in architecture survivability after release, and a lack of high-level sustainability metrics. Future directions include integrating sustainability in education, addressing greenwashing risks, understanding sustainability in software development, and prioritizing sustainability in software design. |
| [25] | 2023 | Factors that influence sustainability aspects in crowdsourced software development: A systematic literature review | The literature review found six critical factors: lack of coding standards in documentation, use of popular programming tools, crowd's lack of knowledge and awareness about sustainability, energy-efficient coding, lack of awareness about sustainable software engineering practices, and lack of coordination/communication between client and crowd. |
| [20] | 2023 | Integrating Sustainability Metrics into Project and Portfolio Performance Assessment in Agile Software Development: A Data-Driven Scoring Model | Despite increasing acknowledgment of sustainability's significance in software engineering and its potential societal and environmental impact, a literature gap remains regarding the integration of sustainability metrics into project and portfolio performance assessment within agile software development contexts. |
| [39] | 2023 | Towards Sustainable Software A Structured Map of Techniques and Best Practices for Organizations and Development Teams to Enhance Sustainability | There is a need for sustainable practices extended to the software industry. |
| [21] | 2023 | A Review of Software Architecture Evaluation Methods for Sustainability Assessment | The results indicate that most of the methods are scenario-based, offering support for technical sustainability but lacking coverage in four-dimensional sustainability. |
| [40] | 2023 | Sustainability is Stratified: Toward a Better Theory of Sustainable Software Engineering | The academic literature on sustainable software engineering lacks empirical studies, needing more evaluations of sustainability interventions. Most papers focus on ecological and product sustainability, rather than processes. The results of the review suggest key propositions: sustainability has different meanings across different levels of abstraction and emerges from interactions from different social, technical, and socio-technical systems. |
| [3] | 2022 | Sustainability in Software Architecture: A Systematic Mapping Study | The results of the review indicate that existing works have primarily focused on specific technical aspects of sustainability, neglecting the holistic perspective needed to address its multi-dimensionality. |
| [48] | 2022 | Models of Sustainable Software: A Scoping Review | This study found that most published models lack verification or validation, especially in software development processes and various aspects of sustainability. None of the models for software development processes were experimentally verified to deliver promised sustainability benefits. Energy efficiency was the most commonly addressed aspect of sustainability in the surveyed models. |

However, the significance of software sustainability extends beyond the development phase. For instance, the proliferation of energy-intensive training models of AI-enabled software systems raises concerns about the ongoing sustainability consideration in software systems. Moreover, several software projects are challenged exceeding budget and resources and others fail, which hinders the sustainability of software systems. These unsuccessful projects are often linked to poor requirements engineering [28]. There is a lack of understanding on how software sustainability should be approached in the requirements engineering phase [4].

Moreover, the absence of tailored approaches for agile settings hampers consensus on sustainable practices. Strategies tested in industrial contexts await broader adoption, posing a challenge in refining these methods for wider practitioner use [4].

Architectural considerations also play a crucial role in sustainable software development. Challenges such as the need for better methods to maintain architectural consistency, the prevalent focus on code-level rather than architectural sustainability, uncertainties in architectural reconstruction, and the lack of overarching sustainability metrics all contribute to the uncertainty [50]. Crowdsourced development, a prevalent practice in software development, faces its own set of sustainability challenges. Factors such as the absence of coding standards in documentation, the prevalent use of popular but energy-intensive programming tools, the lack of knowledge and awareness about sustainable practices among contributors, and insufficient coordination between clients and the crowd all impact sustainability in this sphere [25].

Table 2: Tertiary study results Part II

| Ref. | Year | Title | Challenges or limitations reported in the selected publication |
|---|---|---|---|
| [32] | 2022 | Towards a common understanding of sustainable software development | Recent studies indicate that software engineers lack a common and clear understanding of sustainable software development. Key findings include the following: a presence of technical, environmental, and social concerns in all phases of the software development life cycle, the importance of considering sustainability requirements early in the life cycle, a necessity for stakeholder engagement centered on sustainability, a role of software quality requirements in developing sustainable software, and the potential for trade-offs in projects due to sustainable concerns. |
| [44] | 2021 | The Green Software Measurement Structure Based on Sustainability Perspective | Although sustainability goals can be achieved through balanced dimensions, there is still a lack of measurement tools to ensure the production of environmentally friendly products as needed by industries and societies. |
| [10] | 2021 | Introduction to software sustainability | Various sustainability levels are defined, focused mainly on the environmental dimension. Identified a need for more research on human and economic aspects of software sustainability. |
| [45] | 2021 | Green-Agile Maturity Model: An Evaluation Framework for Global Software Development Vendors | The study identified risk factors found in literature that impact green and sustainable software development when using agile methods in global software development. This includes the following: insufficient system documentation, limited support for real-time and large systems, management overhead, absence of customer presence, lack of formal communication, limited support for reusability, insufficient customer knowledge, and absence of long-term planning. |
| [35] | 2020 | Incorporating sustainability into software projects: a conceptual framework | Most software literature concentrates on either the project product or process, or on a few sustainability dimensions, rather than all three aspects of the triple bottom line theory. This is an accounting framework including three key areas: social, economic and environmental. |
| [42] | 2020 | Sustainable software engineering: A perspective of individual sustainability | The study finds that key challenges for individual sustainability include the following: a lack of domain knowledge, methodologies, tool support, education, diverse and unidentified situations, and established sustainable software engineering practices |
| [29] | 2019 | Towards Green Software Process: A Review on Integration of Sustainability Dimensions and Waste Management | Current studies on software processes often overlook key sustainability dimensions: economic, social, environmental, individual, and technical. The review found that none of the green software process models integrate sustainability or waste reduction in the software development life cycle phase. |
| [30] | 2019 | Quality and sustainability dimensions toward green software product: A review | Previous studies have focused largely on power and energy consumption, waste reduction, and disposal in green hardware. However, there is a noticeable gap in research concerning green software, despite its indirect impact on the environment and the software's sustainability. While some studies exclusively address green software, there is a lack of integration with broader sustainability dimensions. |

### 4.3 Lack of assessment metrics and tools

In software sustainability research, a critical challenge emerges from the lack of comprehensive assessment metrics and tools, a gap that has been emphasized across various reviews [4, 20, 44]. Understanding and evaluating a software system's sustainability necessitates a robust set of metrics and tools, yet this tertiary study reveals an alarming deficiency in this area [1]. Without established metrics, achieving a comprehensive assessment of a software's sustainable impact on various stakeholders, including society, becomes a difficult task. While the essence of sustainability demands a balanced consideration across all dimensions, recent attention has primarily gravitated towards specific metrics that often narrow down to technical aspects and energy consumption [17, 30, 40]. This leaves a significant gap in the evaluation of environmental and socio-economic aspects, highlighting the need for a more holistic approach to assessment metrics.

The absence of standardized frameworks and metrics in software sustainability further compounds the challenge. This lack of uniformity results in varied approaches to sustainability assessment across different software projects and organizations. The consequence is a difficulty in comparing the sustainability performance of different software products or tracking improvements over time. Moreover, the absence of clear frameworks and metrics introduces ambiguity regarding the prioritization of sustainability factors and their measurement methods. This ambiguity can lead to the oversight of crucial metrics, hindering the development of effective sustainability strategies [3, 21, 50]. A standardized framework not only simplifies communication among stakeholders but also fosters more efficient collaboration in sustainable software development efforts.

A significant challenge in achieving comprehensive sustainability metrics lies in the specialized knowledge required for each software sustainability dimension. Developing frameworks and metrics that adequately capture all facets of sustainability necessitates interdisciplinary collaboration and a holistic approach to software evaluation. In response to these challenges, Guldner et al. have introduced a Green Software Measurement Model, building upon previous models such as the GREENSOFT model, to focus on energy evaluations [24, 33]. This model serves as a step towards a more comprehensive and standardized approach to sustainability measurement in software systems.

Despite these efforts, a notable literature gap remains in integrating sustainability metrics into project and portfolio performance assessments within agile software development contexts [20]. Moreover, the current academic discourse on sustainable software engineering calls for more empirical studies, particularly in evaluating software sustainability solution proposals. The predominant focus on ecological and product sustainability highlights the need for a broader consideration of sustainable processes within software development [40].

### 4.4 Narrow perspectives on sustainability in software systems

The current research in sustainable software systems is characterized by its narrow focus, often fixating on a specific dimension [41], or an isolated phase of software development [47]. This limited perspective overlooks the multi-dimensionality inherent in sustainable software, failing to address the broader contexts that extend beyond the software itself. While numerous research studies have proposed solutions and approaches for sustainable software systems [8, 41, 50], they often fall short in their evaluation and tend to narrow their focus on singular aspects, such as energy efficiency or the durability of the software product [40, 41, 48]. This narrow focus risks impeding the overarching goal of achieving sustainable software systems.

Various models proposed for sustainable software development have been found to be deficient in addressing all sustainability dimensions or are lacking in their coverage of specific software phases [3, 21, 29, 32, 47]. For instance, one study focused on the individual dimension of sustainability, revealing its unclear nature within the context of software systems [42]. Another literature review highlighted the scarcity of work on the individual and economic aspects of sustainability within software development [10]. In sustainable software architecture evaluation research, the prevalent methods are often scenario-based, providing support for technical sustainability while lacking comprehensive coverage in all four dimensions of sustainability [21]. This limited approach poses challenges as it fails to consider the broader social, economic, and environmental impacts of software systems.

A crucial consideration in understanding the breadth of sustainability in software systems is the alignment with the United Nations' 17 sustainable development goals (SDGs). However, the connection between software quality standards, such as ISO/IEC 25010, and the comprehensive scope of sustainability goals remains largely unexplored [31]. This lack of alignment not only limits the potential impact of software systems on broader societal goals but also underscores the need for a more holistic approach to sustainable software development.

By fixating on singular dimensions or aspects of sustainability, there is a risk of neglecting the holistic perspective needed for true software sustainability. Focusing solely on technical efficiency, for example, may inadvertently lead to software that lacks usability or fails to meet the diverse needs of its users. Sustainable software development necessitates a comprehensive approach that considers the well-being of users, the broader societal impacts, and the intricate interplay between various dimensions of sustainability. Without this holistic view, software systems may fall short of their potential to contribute meaningfully to a sustainable future.

### 4.5 Lack of awareness

The awareness deficit surrounding sustainable software spans across multiple fronts, encompassing education, industry practices, and end-user understanding. This deficiency in awareness can be attributed to several factors, including the inherent complexity of sustainability concepts, limited education and training opportunities, and the challenge of making software sustainability concepts accessible to a broader audience with varying levels of technical expertise [43]. In industry, a deficiency in established methodologies for incorporating sustainability persists [4, 25, 39]. Notably, within the context of crowdsourcing, where awareness gaps are pronounced, the integration of sustainability practices remains an unaddressed challenge [25]. Similarly, the incorporation of sustainability within agile methodologies has received limited attention, indicating a need for further exploration in this domain [4, 20, 45].

Unlike physical products, where the environmental impact may be more visible, the sustainability of software often operates discreetly in the background, escaping the notice of end users. This lack of immediate visibility can result in users prioritizing factors such as functionality, cost, or convenience over sustainability considerations when interacting with software. Without a deliberate emphasis on integrating sustainability into software design and development, end users may remain unaware of its significance. Education is a pivotal point of intervention, particularly in terms of software sustainability for IT professionals [42, 50].

Currently, the field lacks a clear methodology or standardized coverage of sustainability in educational curricula. The ongoing research on sustainability in education underscores the critical need to integrate these concepts into learning environments. Without adequate education on sustainability, future developers and designers may lack the necessary skill sets to contribute meaningfully to sustainable practices in industry and society. This deficit not only impacts the potential for sustainable software development, but also hinders the understanding of individual actions' impact on sustainability goals.

### 4.6 Lack of serious considerations in practice

Within the ICT sector, a sense of skepticism looms large regarding the practical impact of sustainability efforts [50]. A recent survey in 2022 revealed that a staggering 91% of 550 technology executives across 11 countries view actions towards software sustainability as merely greenwashing, lacking substantive effects [37]. The concept of greenwashing, as categorized by Chen and Chang [13], involves products that mislead with ambiguous environmental claims, overstate green functionality, or mask important information to appear more environmentally friendly than they are. Terms such as "sustainable software" and "green software" are sometimes dismissed as fleeting trends rather than serious considerations.

This skepticism often arises due to the absence of clear metrics that substantiate claims of sustainability. The lack of established benchmarks and standards in software sustainability exacerbates this issue, leaving room for ambiguous definitions and interpretations [50]. Moreover, recent developments such as the European Union (EU) parliament's law combating greenwashing, underscore the urgent need for reliable approaches to define and measure the overall sustainability of software products and systems. Without a concrete certification or recognized standards, proving the sustainability of software becomes challenging, if not impossible. Moreover, the dismissal of sustainability as a serious consideration can also be attributed to the limited awareness among developers and stakeholders regarding the significance of sustainable software. The intricate socio-technical challenges associated with addressing sustainability further complicate matters [50].

## 5 RECOMMENDATIONS

Based on the results of this tertiary study, we propose the following recommendations to clear the path for software sustainability.

### 5.1 Need for a comprehensive approach

To establish sustainable software, it is imperative to adopt a balanced approach with comprehensive metrics for assessment. A crucial step forward is to broaden the research focus, moving beyond narrow scopes such as environmental or energy-centric viewpoints. Exploring sustainability throughout various phases of software development, including requirements engineering, architecture, development processes, and metrics evaluation, is essential for comprehensive solutions [47, 50]. Promoting awareness and education on sustainable software engineering practices, particularly at the university level, is critical. There is a need to address the existing lack of knowledge and awareness among software engineers, especially in dynamically evolving fields [25].

Comprehensive evaluation methods are pivotal for assessing the sustainability of software architectures. These methods should encompass not only technical sustainability but also broader dimensions such as social, economic, and environmental factors. Moving away from scenario-based assessments towards holistic approaches ensures a more accurate reflection of software sustainability [21]. As highlighted by Guldner et al. [24], the process of identifying models and tools should be iterative, involving diverse stakeholders to gain a thorough understanding of the problem and develop effective sustainable software solutions. Encouraging a holistic perspective in software architecture for sustainability is key to effective integration. This approach entails considering not just of the technical facets but also the social, economic, and environmental implications throughout the software design and development processes [3].

### 5.2 Sustainability as the foundation of software development

To establish a foundation of sustainability in software development, we highlight the need for a comprehensive approach that begins from the project's inception and extends throughout its entire lifecycle. This involves incorporating society and nature as key stakeholders, addressing the multifaceted dimensions of social, environmental, and economic sustainability. A clear definition of sustainable software is crucial for guiding development efforts. We agree with the definition of sustainable software as software that seeks to minimize its environmental impact and maximize resource efficiency across all stages of its lifecycle, while also taking into account economic and social considerations [10, 39].

To promote sustainable practices on an industry-wide scale, it is essential to encourage the adoption of sustainable techniques and frameworks across organizations and development teams [39]. This broad adoption ensures that sustainability is not just an afterthought but an integral part of the software development process from the outset. Integrating sustainability metrics into software development practices is another crucial step. This includes closing the existing literature gap by incorporating sustainability metrics into project and portfolio performance assessments, especially within agile software development contexts. Developing data-driven scoring models and tools can facilitate effective evaluation of sustainability aspects within software projects [20].

### 5.3 Need for actionable research

The need for actionable research in sustainable software practices is evident in the software industry's growing focus on sustainability [19, 39]. By focusing on proposing actionable research, the software industry can move towards more sustainable practices, ensuring that software development not only meets technical requirements but also aligns with broader environmental, economic, and social sustainability goals. One key area that requires attention is the integration of sustainability into agile software development processes. Tailored approaches, developed through consensus-based strategies and tools, are essential to effectively incorporate sustainability into the fast-paced environment of agile development [4].

Similar to the multifaceted nature of quality in software, sustainability encompasses various dimensions that require clear guidelines and metrics for effective management. Just as metrics such as code coverage and mean time between failures are used to assess software quality, there is a need for standardized frameworks and metrics to evaluate software sustainability. Empirical studies play a crucial role in advancing our understanding of sustainable software engineering. It is imperative to evaluate the effectiveness of sustainability interventions and explore sustainability from various perspectives, including environmental, social, technical, and economic dimensions [40]. Moreover, models proposed for sustainable software development must undergo rigorous validation and verification processes. Experimental testing is essential to confirm the promised sustainability benefits, especially within the context of software development processes [48].

### 5.4 Need for a common understanding

The concept of software sustainability goes beyond individual projects or organizations; it holds global implications, affecting the environment, society, and economy in multifaceted ways. To establish a common understanding of sustainable software, it is crucial to address the unnecessary confusion caused by terms like "sustainable-by" and "green-by." When we aim to make a software system sustainable, the focus should be on adding sustainability to the system itself, rather than labeling it as *sustainable software*. In software sustainability there exist first, second, and third order effects, which pertain to the impact of using software on sustainability dimensions. These effects should be thoroughly understood and integrated into discussions around sustainability metrics and frameworks.

Promoting awareness and education is another crucial aspect of fostering a shared understanding of sustainable software engineering practices. Increasing awareness among software engineers is critical, for instance in crowdsourced development settings [25]. Moreover, considering the prevalence of agile methodology in the workplace, there is a need for further research to better understand how to integrate software sustainability into agile practices [20]. One effective method to raise awareness of software sustainability could be through the use of the Sustainability Awareness Framework (SusAF). This framework has undergone multiple iterations and has been successfully used and evaluated in both industry and educational settings [7, 18].

### 5.5 Need for diverse perspectives

Software holds the potential to shape society, making it imperative to develop solutions that are inclusive, equitable, and beneficial for all users, regardless of their backgrounds or abilities. Within the software sustainability complex ecosystem, designers, users, creators, and the environment all play crucial roles as stakeholders. To truly understand and address software sustainability, a diverse range of perspectives is essential. The global and multi-layered nature of software sustainability demands the inclusion of varied opinions and viewpoints [32]. Diverse perspectives enrich the discourse, helping to uncover blind spots and guiding the integration of software sustainability into global, business, and user needs. Fostering inclusive decision-making processes paves the way for innovative solutions that not only address the socio-economic impact of software but also minimize its environmental footprint.

Considering diverse perspectives becomes crucial when balancing global sustainability goals, business objectives, and user needs. For instance, global sustainability aims may focus on reducing carbon emissions and promoting equitable access to technology, while business goals revolve around profitability and market competitiveness. User needs, on the other hand, span from usability and functionality to ethical considerations and data privacy. The inclusion of a wide range of perspectives enriches discussions surrounding software sustainability. It encourages exploration of best practices, innovative solutions, and ethical considerations. Without diverse viewpoints, there is a risk of overlooking critical aspects of software sustainability, such as social inequalities, environmental risks, or economic inefficiencies. Embracing diversity in perspectives ensures a more holistic and effective approach towards sustainable software development.

### 6 LIMITATIONS

This paper may have some limitations, such as: (i) The search for candidate papers was conducted exclusively in Scopus. While this may seem restrictive, many papers from other digital libraries like IEEE, ACM, and Springer are also available in Scopus. Including studies from Scopus could also serve as a quality filter since Scopus indexes peer-reviewed papers from publication venues that meet certain criteria, such as regular publishing and having expert editors. (ii) Terms related to the topic might have been missing from the search string. However, the current search resulted in a good number of relevant studies on software sustainability published in the last five years. (iii) Part of the search string focused primarily on titles in Scopus to ensure relevance, potentially overlooking studies that did not use these specified terms. However, we assume that papers addressing software sustainability would generally include these terms in their titles.

### 7 CONCLUSION

The challenges outlined in this tertiary study, such as the ambiguous definition of software sustainability, and gaps in education, assessment, and awareness, underscore the complexities of integrating sustainability into software engineering practices. Given the expansive and evolving nature of sustainability, further research—particularly empirical studies in industry—can deepen our understanding of its dimensions and evaluation methods, thereby advancing sustainable practices in software engineering. Addressing these challenges will necessitate collaborative efforts from academia, industry, policymakers, and society at large. Establishing clear standards, promoting education on software sustainability, and increasing end-user awareness are pivotal steps toward a future where sustainability is not merely a concept but an integral component of every software system.